\documentclass[twocolumn,preprintnumbers,amsmath,amssymb]{revtex4}
\usepackage{graphicx,color}
\usepackage{amsmath,amssymb,latexsym}

\begin{document}

\title{Finite-size scaling in unbiased translocation dynamics}

\author{Giovanni Brandani}
\email{s1246659@sms.ed.ac.uk}
\affiliation{
University of Edinburgh, SUPA, School of Physics, Mayfield Road, EH9 3JZ, UK
}

\author{Fulvio Baldovin}
\email{baldovin@pd.infn.it}
\affiliation{
Dipartimento di Fisica e Astronomia and Sezione INFN,
Universit\`a di Padova,
 Via Marzolo 8, I-35131 Padova, Italy
}

\author{Enzo Orlandini}
\email{orlandini@pd.infn.it}
\affiliation{
Dipartimento di Fisica e Astronomia and Sezione INFN,
Universit\`a di Padova,
 Via Marzolo 8, I-35131 Padova, Italy
}

\author{Attilio L. Stella}
\email{stella@pd.infn.it}
\affiliation{
Dipartimento di Fisica e Astronomia and Sezione INFN, 
Universit\`a di Padova,
 Via Marzolo 8, I-35131 Padova, Italy
}

\date{\today}

\begin{abstract}
Finite-size scaling arguments naturally lead us to introduce a
coordinate-dependent diffusion coefficient in a Fokker-Planck
description of the late stage dynamics of unbiased polymer
translocation through a membrane pore.
The solution for the probability density function of the chemical
coordinate matches the initial-stage subdiffusive regime and takes
into account the equilibrium entropic drive. 
Precise scaling relations connect the subdiffusion exponent to the
divergence with the polymer length of the translocation time,
and also to
the singularity of the probability density function at the absorbing
boundaries.
Quantitative comparisons with numerical simulation data in $d=2$  
strongly support the validity of the
model and of the predicted scalings.
\end{abstract}

\maketitle

Translocation of long polymers across a membrane is a  basic
biological process~\cite{Kasianoviz:1996:PNAS,Meller_et_al:2001:PRL,Dekker_2007_Nature_nano} 
and a fundamental problem in polymer dynamics.
In the last fifteen years, 
a number of facts have been established and open issues pointed out 
\cite{Sung1,Muthukumar1,Kardar1,Metzler3,Ali,Barkema1,Barkema2,Dubbeldam1,Kardar4,Luo2,Zoia1,Barkema7}. 
Early attempts~\cite{Sung1,Muthukumar1} to study unbiased translocation on the basis of
quasi-equilibrium assumptions and Fokker-Planck equations for the chemical-coordinate 
probability density function (PDF) revealed inadequate for long chains
\cite{Kardar1}. 
There is now consensus~\cite{Kardar1,Kardar4,Barkema1}
on the fact, that during the initial stages, the 
scaling with time of the mean square displacement  
is sub-diffusive, and clear evidence points out that within this anomalous stage  the process is
well reproduced by a fractional Brownian motion.
The precise value of the sub-diffusive exponent still remains
controversial~\cite{Barkema2,Kardar1,Kardar4,Luo2,Barkema7}
and could even depend on the viscosity of the solvent~\cite{Slater2}.  
Most important, for finite lentgths of the chain 
(which is a necessary condition for the translocation
process to occur) the fractional Brownian motion
description breaks down at times in which the translocation process
has not occurred yet~\cite{Barkema1,Barkema7}. 
The result is that, to the best of our knowledge, 
no theory is presently capable of quantitatively reproducing neither the
chemical-coordinate PDF, nor the survival probability in the whole
time span of the process.
It is interesting to notice that, in spite of the breaking down of the initial subdiffusion behavior, 
signatures of it could remain in the late stages of the process; 
The singular behavior
of the asymptotic PDFs of polymer displacements at the boundary
values of the translocation coordinate could be such a candidate~\cite{Kardar4}. 
It is not clear however how these singularities can be linked to the initial anomalous
diffusion regime. 

Here we show that the unbiased translocation process, in the 
time window following the initial anomalous diffusion, is in fact
described by a Fokker-Planck equation with a displacement-dependent
diffusion coefficient which neatly originates from a finite-size
scaling analysis. 
Based on only two free parameters associated with microscopic details, 
the Fokker-Planck equation quantitatively
reproduces numerical results for 
both the translocation
coordinate PDF and the survival probability. 
In addition, the finite-size scaling properties of the diffusion
coefficient provide the mechanism for the appeareance of singularities
in the long-time PDF of the
translocation regime:
For the first time we are thus able to furnish a theoretical explanation
of these singularities by linking them to the anomalous 
scaling exponent of the initial stages of the process.

\begin{figure}
\includegraphics[width=1.0\columnwidth]{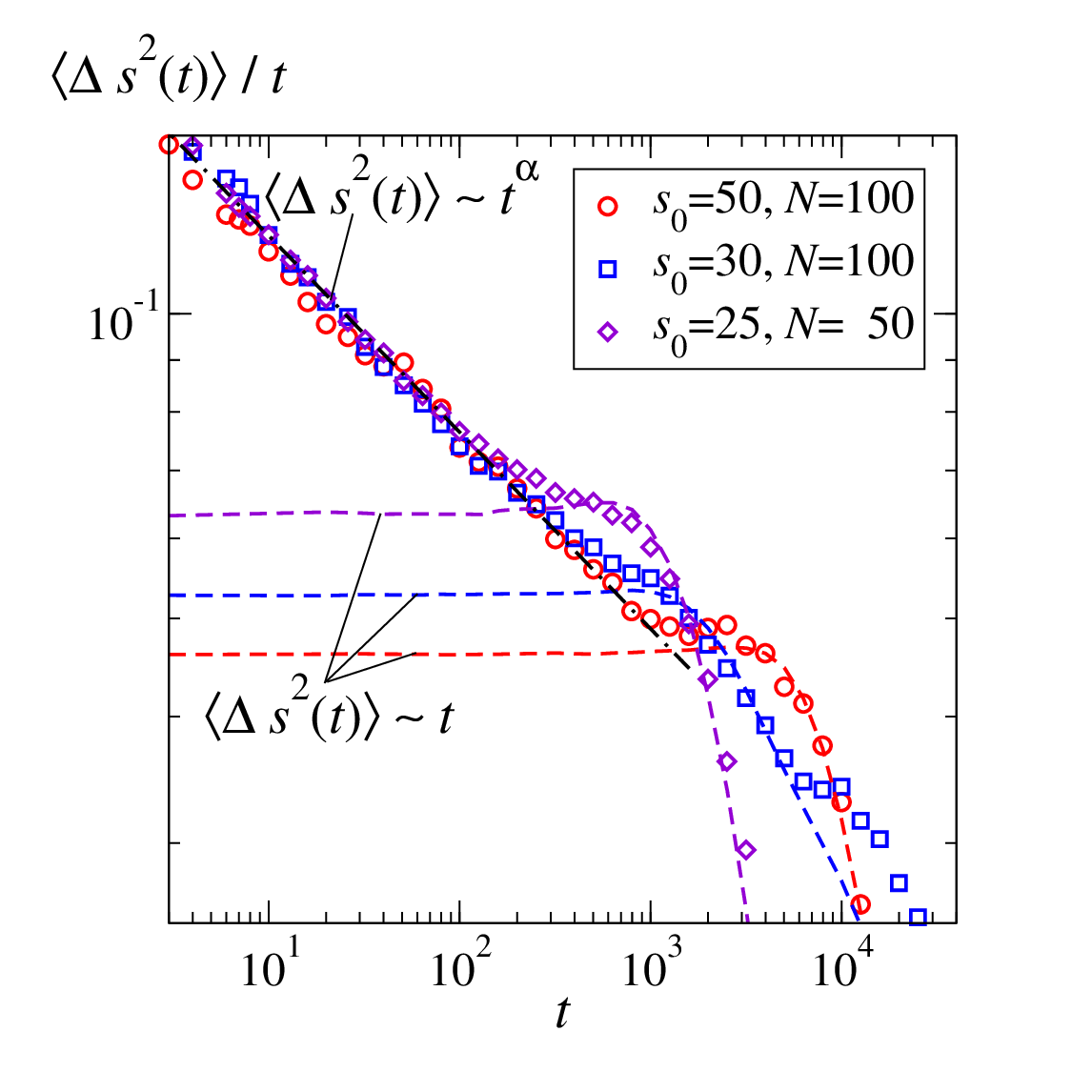}\\
\caption{
  Symbols: Time-rescaled mean square displacement evolution for a translocating
  chain. At early stages, the dynamics follows a
  universal sub-diffusive behavior
  $\langle\Delta s^2(t)\rangle\sim t^\alpha$, with
  $\alpha\simeq0.81$ (dot-dashed line).
  This regime breaks-down at a time $\tau$ depending on
  both $s_0$ and $N$. The dashed curves 
  are obtained by numerically solving Eq. (\ref{eq_fp_1}) 
  with $D(s,N)$ given by Eq. (\ref{eq_d_s0}), $A=0.13$, and $s(0)=s_0$. 
}
\label{fig_msd_langevin}
\end{figure}

Our numerical results are based on molecular dynamics simulations 
for the translocation dynamics of 2d self-avoiding linear chains made by $N+1$ monomers
of unit length and described by a FENE-Shifted Lennard-Jones interaction potential, in
contact with a Langevin heat bath. 
By indicating with $0\leq s(t)\leq N$ the 
number of monomers at one side of the pore at time $t$,
the symbols in Fig. \ref{fig_msd_langevin} refer to the (time-rescaled) mean square displacement 
$\langle\Delta s^2(t)\rangle\equiv\langle [s(t)-s_0]^2\rangle$ for
chains initially equilibrated with the monomer $s(0)=s_0$ at the pore.
The first important feature to note in Fig. \ref{fig_msd_langevin} is the existence of an
initial anomalous regime, independent of both $N$ and
$s_0$, during which $\langle\Delta s^2(t)\rangle=2\,D_\alpha\,t^\alpha$ with
$\alpha\simeq0.81$ and $D_\alpha$ a generalized diffusion
coefficient. 
The precise value of $\alpha$, 
which is bound from below by $(1+\nu)/(1+2\nu)$~\cite{Barkema2} 
and from above by $2/(1+2\nu)$~\cite{Kardar1},
is still under debate~\cite{Barkema2,Kardar1,Kardar4,Luo2,Barkema7}; 
Recently, it has been claimed that $\alpha$ is also viscosity-dependent \cite{Slater2}. 
For the system sizes analyzed here, we have found values of $\alpha$
close to the upper 
bound pointed out in~\cite{Kardar1}.
In any case, our approach does not imply or require universality of
this exponent. 
This $N$ and $s_0$ independent initial
stage can be ascribed to the scale-free, self-similar structure of the polymer,
which is explored by the translocating coordinate. 
The finite size of the polymer implies a break-down of the initial regime, 
which is followed by a stage in which the growth of 
$\langle\Delta s^2(t)\rangle$ is closer to linear in $t$, 
before dropping down as a consequence of the fact that 
with finite probability the translocation process has been completed. 
The times $\tau$ at which the break-down of the initial anomalous
scaling occurs, of course depend on both $s_0$ and $N$: 
$\tau=\tau(s_0,N)$~\cite{foot1}.
Thus, if one wishes to {\it match} the behavior of 
$\langle\Delta s^2(t)\rangle$ just after the break-down 
with that of a normal diffusion starting at $t=0$ with 
$s=s_0$, the normal diffusion coefficient $D(s_0,N)$
should satisfy the condition
\begin{equation}
\label{eq_crossover_fp}
2\;D_{\alpha}\;\tau^\alpha(s_0,N)
=2\;D(s_0,N)\;\tau(s_0,N).
\end{equation}
Below, we argue that the idea of matching the evolution of the whole
process for $t>\tau$ with an effective Fokker-Planck description
indeed works very well, provided one promotes the diffusion
coefficient $D(s_0,N)$ identified by Eq. (\ref{eq_crossover_fp})
to enter, with its coordinate dependence, in the Fokker-Planck equation. 

Treating $s$ as a continuous coordinate, our goal is thus to develop an
effective Fokker-Planck equation to be satisfied by the PDF $p(s,t|s_0,N)$
of having the monomer $s$ at the pore at time $t$, given that the
translocation process, for a polymer of size $N$, started at time zero
with $s_0$ monomers at one side of the wall. The problem has natural
absorbing boundaries at $s=0$ and $s=N$, corresponding to the configurations
in which the polymer completes the translocation process.  
We wish to profit of Eq. (\ref{eq_crossover_fp}) and 
start to construct our model by first neglecting the entropic drive on the
process. 
According to It\^o's rules for stochastic integration~\cite{foot2}, 
an over-damped Fokker-Planck equation with a position- 
and size-dependent diffusion coefficient, $D(s,N)$, reads
\begin{equation}
\label{eq_fp_1}
\partial_t\;p(s,t|s_0,N)=\partial^2_s\left[D(s,N)\;p(s,t|s_0,N)\right].
\end{equation}
Multiplying Eq. (\ref{eq_fp_1}) by $(s-s_0)^2$ and integrating over
$s$, one easily obtains
\begin{eqnarray}
\partial_t\mathbb E\left[(s-s_0)^2\right]
&=&\int_0^N d s\;2\;D(s,N)\;p(s,t|s_0,N)
\nonumber\\
\label{eq_approximation}
&\simeq& 2\;D(s_0,N),
\end{eqnarray}
where the approximation holds for times at which $p(s,t|s_0,N)$ is
sufficiently concentrated around $s_0$ or for $D(s_0,N)$ slowly
varying in such range.
Below, we confirm {\it a posteriori} 
that these conditions are verified as long as the survival probability
$S(t|s_0,N)\equiv\int_0^N d s\;p(s,t|s_0,N)$
is close to 1.
Within this approximation, one is entitled to identify
in Fig. \ref{fig_msd_langevin} the ordinates at $t=\tau(s_0,N)$ 
as $2\;D(s_0,N)$. 

\begin{figure}
\includegraphics[width=1.0\columnwidth]{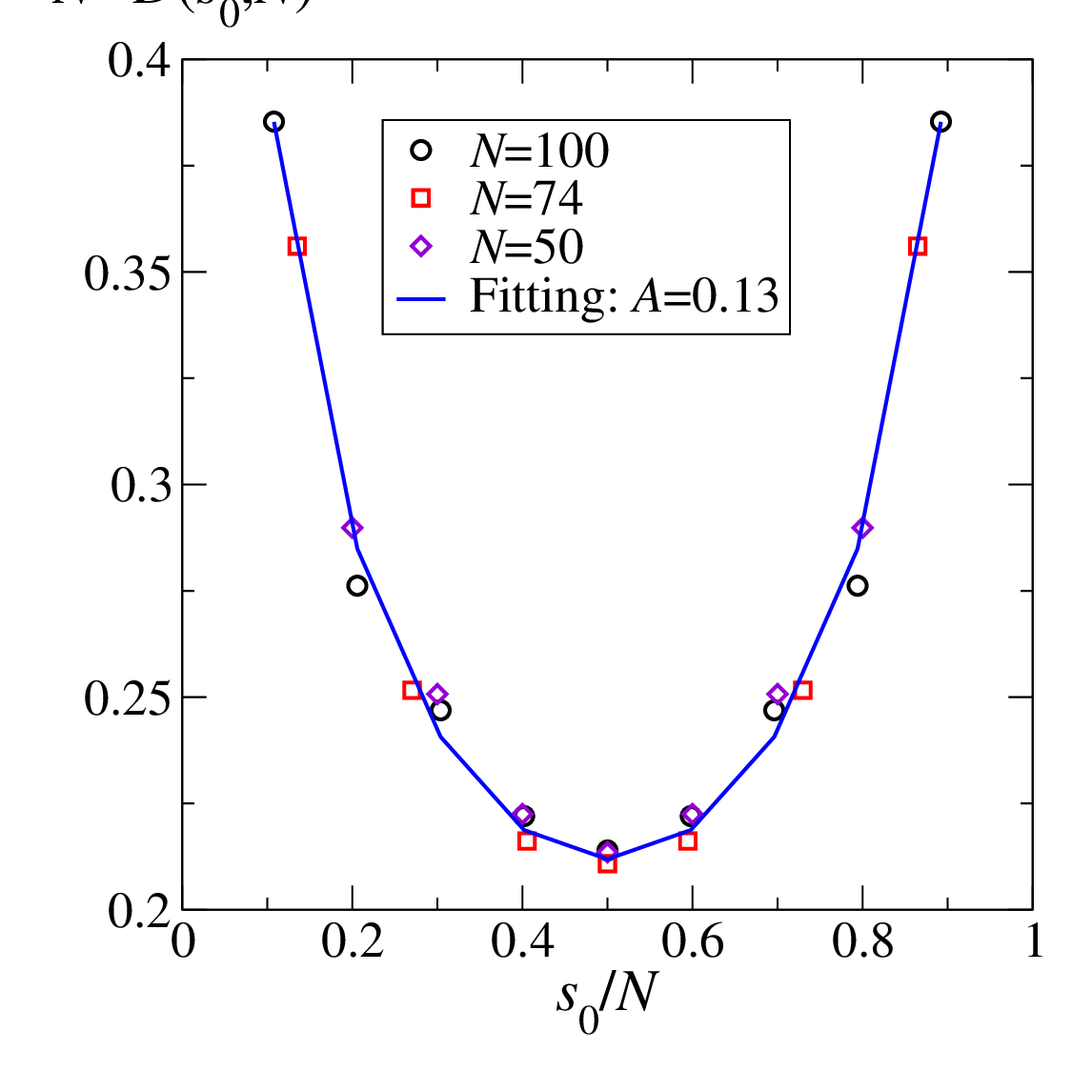}\\
\caption{
  Data-collapse of $D(s_0,N)$ according to Eq. (\ref{eq_d_s0}). 
  Points are obtained from translocation simulations 
  by averaging 
  $\langle\Delta s^2(t)\rangle/2t$ between $\tau(s_0,N)$ and the
  maximum time for which $S(t|s_0,N)\simeq1$.
}
\label{fig_d_s}
\end{figure}

By Eq. (\ref{eq_crossover_fp}) one gets
$D(s_0,N)\propto\tau(s_0,N)^{\alpha-1}$~\cite{foot1}.
For Rouse dynamics we have $\tau(N/2,N)\sim N^{1+2\nu}$
($\nu=3/4$ for 2d self-avoiding polymers~\cite{Nienhuis}).
For $s_0\neq N/2$ we expect 
$\tau(s_0,N)$ to obey a scaling law both in $s_0$ and $N$ of the 
form
$
\tau(s_0,N)\sim N^{1+2\nu}\;f\left(\frac{s_0}{N}\right)
$
with $f(x)=f(1-x)$, or
\begin{equation}
D(s_0,N)\propto
\;\left[
\frac{1}{N^{1+2\nu}}
\;\frac{1}{f\left(\frac{s_0}{N}\right)}
\right]^{1-\alpha}.
\end{equation}
A further condition can be found for $s_0\ll N$.
In this case, only the branch of the chain with length  $s_0$
breaks the self-similarity sustaining the anomalous scaling (since
the other branch becomes arbitrarily long). 
Thus, 
$
\tau(s_0,N)\;\substack{\sim \\ s_o\ll N}\;s_0^{1+2\nu}
$,
implying
$
f(x)\;\substack{\sim \\ x\ll 1}\;x^{1+2\nu}.
$
By putting together this small-argument behavior with the symmetry $f(x)=f(1-x)$, 
it is reasonable to guess
\begin{equation}
\label{eq_d_s0}
D(s_0,N)
=\frac{A}{N^\sigma}
\;\left[
\frac{1}{\left(\frac{s_0}{N}\right)^{1+2\nu}}
+\frac{1}{\left(1-\frac{s_0}{N}\right)^{1+2\nu}}
\right]^{1-\alpha},
\end{equation}
where $A$ is a
size-independent coefficient characterized by the
microscopic details of the polymer's dynamics
and $\sigma\equiv(1+2\nu)\;(1-\alpha)$. 
Notice that the maximum crossover time is associated to the central monomer,
$\tau(N/2,N)\sim N^{(1+2\nu)}$.
On the other hand,  as $s_0\to0$ (or $s_0\to N$), $\tau(s_0,N) \to 0$ 
and correspondingly $D(s_0,N)$ in Eq. (\ref{eq_d_s0}) diverges.
We will see below that these divergences,
which are hard to directly detect numerically, generate a singular behavior
at the borders of $p(s,t|s_0,N)$.  

As a first important check of the above deduction, we 
verify the finite-size scaling implied by Eq. (\ref{eq_d_s0}).
In Fig. \ref{fig_d_s} we data-collapse
the numerical results of translocation dynamics obtained with various
$N$ and $s_0$. 
Confirmation of Eq. (\ref{eq_d_s0}) is remarkable,
yielding a best fitted $A\simeq0.13$.
Further validations of our approach are furnished by the comparison
of the solution of Eq. (\ref{eq_fp_1}) with the molecular dynamics
simulations of the translocation process, 
as reported in Fig. \ref{fig_msd_langevin} and specifically commented 
in the Supplemental Material. 
Such a comparison also points out that our theory is expected to work
well for $t>\tau(N/2,N)$ and $s_0=N/2$, and to become less accurate as $s_0$ moves closer
to the ends of the polymer chain.

Similarly to what has been done in the original approach 
by Sung and Park \cite{Sung1}, 
Eq. (\ref{eq_fp_1}) can be improved by considering the effect of an
entropic force
$-\frac{D(s,N)}{k_B T}\frac{d F(s,N)}{d s}$,
where $F(s,N)=-k_B T\;\ln\Omega(s,N)$ is the free energy of the constrained polymer, and the
validity of the Einstein relation has been assumed. 
Standard results~\cite{Vanderzande:1998,Madras1,Duplantier1}
for the number of possible polymer configurations with the
monomer $s$ at the pore, $\Omega(s,N)$, including scaling corrections, 
yield
\begin{equation}
\label{eq_omega}
\Omega(s,N)
\propto
[s\;(N-s)]^{(\gamma_1-1)}
\;C(s)\;C(N-s),
\end{equation}
where 
the surface entropic critical exponent is 
$\gamma_1=61/64=0.95$ for 2d linear polymers with self-avoidance
~\cite{Duplantier1}, 
and
\begin{equation}
\label{eq_correction}
C(s)\simeq1+\frac{b_0}{s^{1/2}},
\end{equation}
with $b_0$ a parameter depending on the microscopic details of
the model. 
Indeed, the inclusion of the entropic drive with a best-fitted
$b_0=0.4$ (with $T=1.2$ in natural dimensionless units) 
improves by about $5\%$ our overall results. 
With the limitations mentioned in the previous paragraph, 
our complete description is thus given by the solution of the 
following equation: 
\begin{eqnarray}
\label{eq_fp_2}
\partial_t\;p(s,t|s_0,N)&=&
-\partial_s\left\{D(s,N)\;\frac{\partial_s\ln\Omega(s,N)}{\partial s}\;p(s,t|s_0,N)\right\}
\nonumber\\
&&+\partial^2_s\left[D(s,N)\;p(s,t|s_0,N)\right],
\end{eqnarray}
with $D(s,N)$ as in Eq. (\ref{eq_d_s0}), $\Omega(s,N)$ as in
Eqs. (\ref{eq_omega}), (\ref{eq_correction}), 
initial conditions 
$p(s,t|s_0,N)=\delta(s-s_0)$, and absorbing boundaries at $s=0$ and $s=N$.

\begin{figure}
\includegraphics[width=1.0\columnwidth]{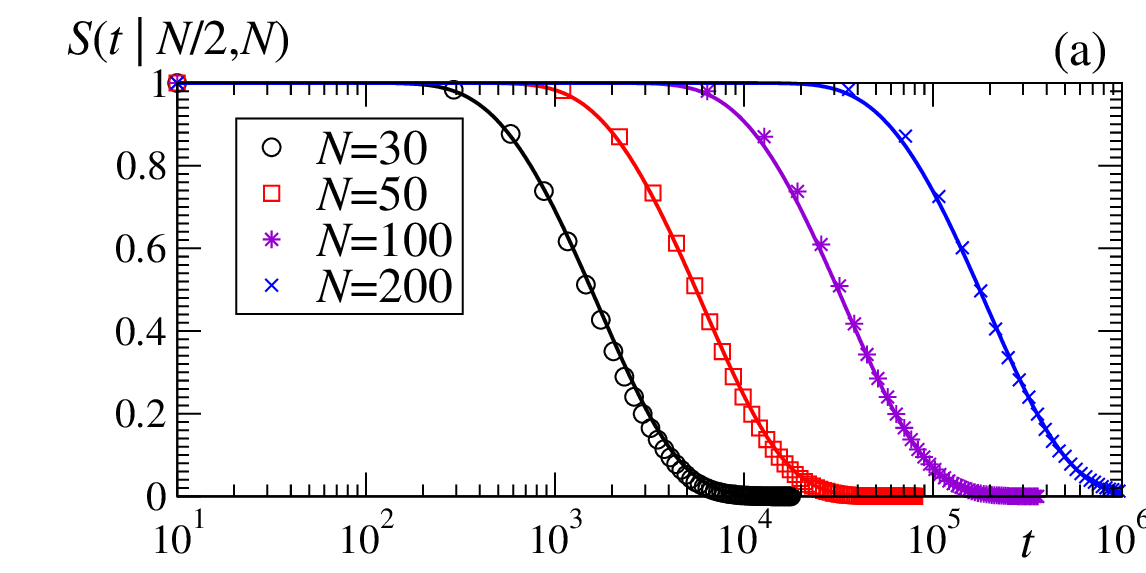}\\
\smallskip
\includegraphics[width=1.0\columnwidth]{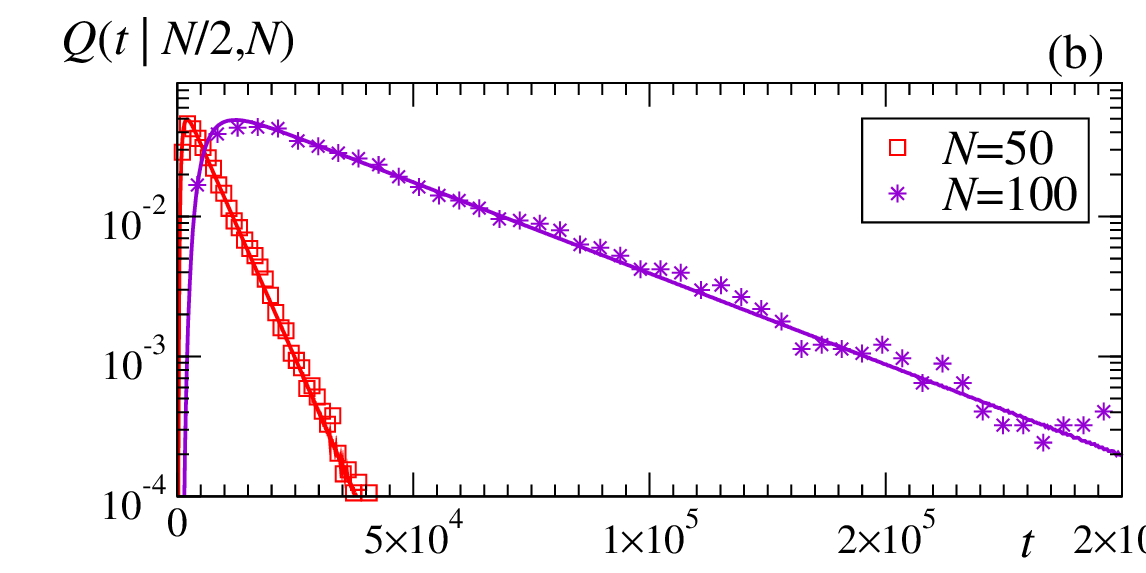}
\caption{Survival (a) and translocation (b) probability 
  for 2d linear polymers: MD simulations (symbols) vs. theory (full lines).}
\label{fig_survival}
\end{figure}

\begin{figure}
\includegraphics[width=1.0\columnwidth]{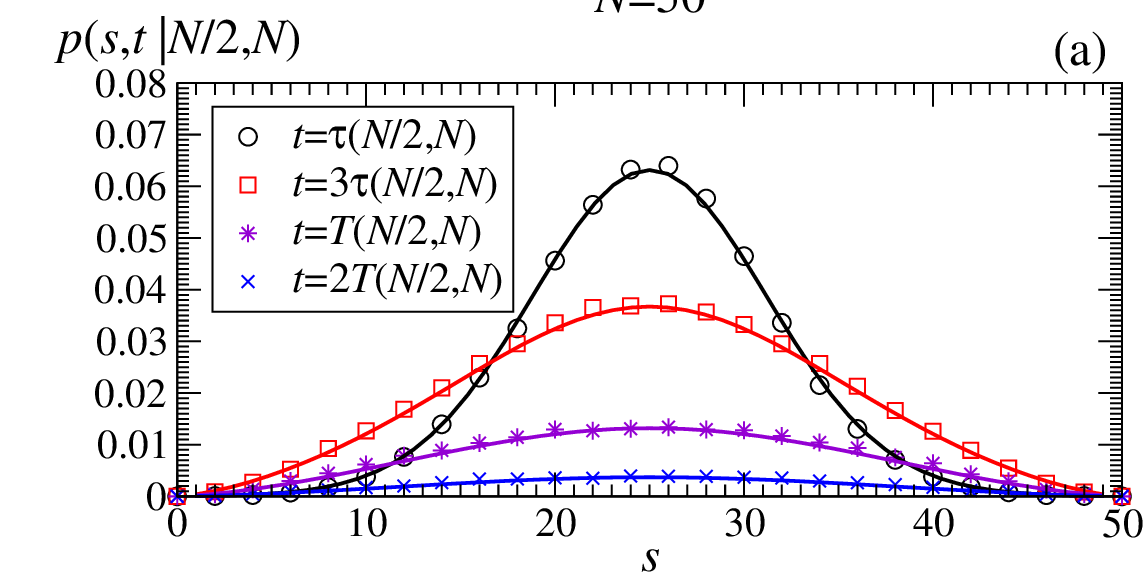}\\
\smallskip
\includegraphics[width=1.0\columnwidth]{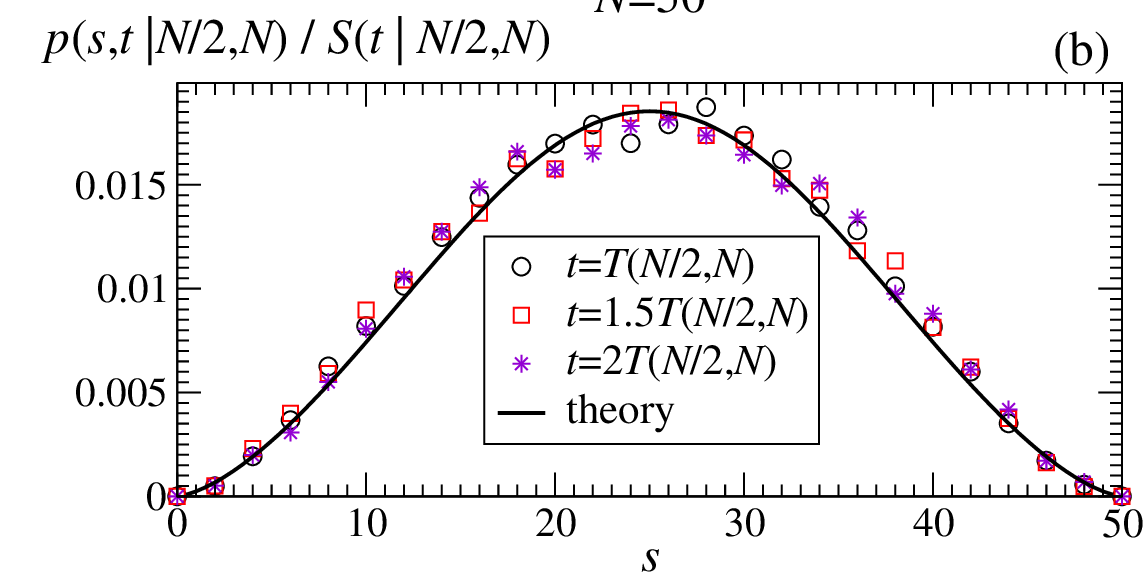}\\
\smallskip
\includegraphics[width=1.0\columnwidth]{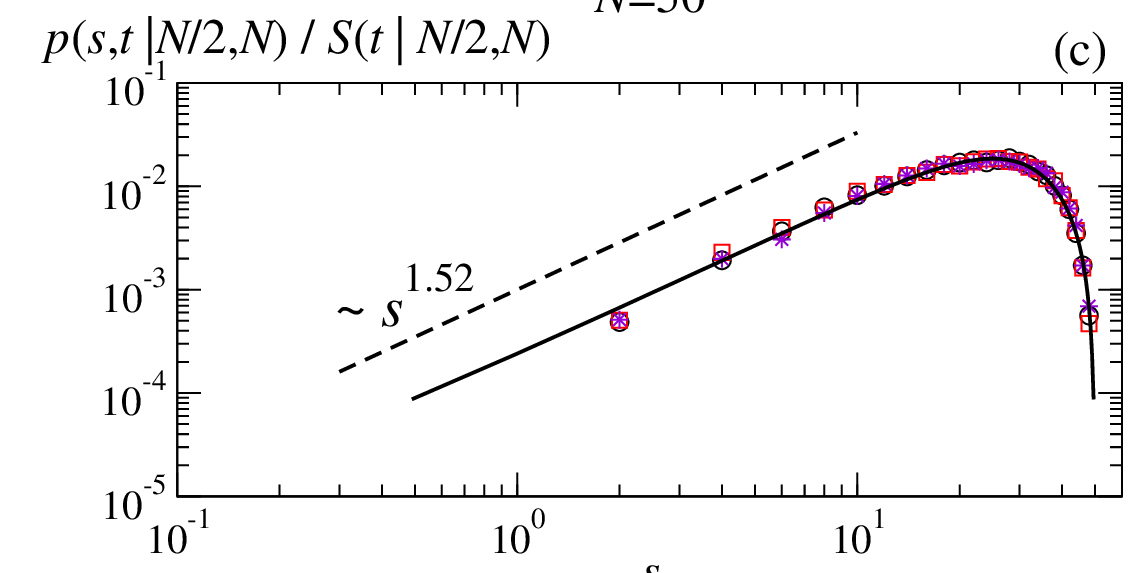}
\caption{Time evolution of $p(s,t|N/2,N)$ (a) and PDF of the surviving
  polymers $p(s,t|N/2,N)/S(t|N/2,N)$ for large time (b): 
  MD simulations (symbols) vs. theory (full lines).
  (c) Same as (b) in log-log scale, to appreciate the singular behavior 
  at the boundary. 
}
\label{fig_pdf}
\end{figure}

Eq. (\ref{eq_fp_2}) is
separable, so that its general solution can be written in the form
\begin{equation}
p(s,t|s_0,N)=\sum_{m=1}^\infty
A_m(s_0,N)
\;X_m(s,N)
\;\mathrm{e}^{-\lambda_m^2\;t},
\end{equation}
where $X_m$ are the eigenfunction corresponding to the 
eigenvalues $\lambda_m$ ($0<\lambda_1<\lambda_2<\cdots$), and the
amplitudes $A_m(s_0,N)$ are determined by the initial conditions. 
For large time, the behavior of the solution is dominated by the
smallest eigenvalue $\lambda_1$. This implies that
both the survival probability $S(t|s_0,N)$ and the translocation
probability $Q(t|s_0,N)=-\partial_t S(t|s_0,N)$ decay exponentially
in the long-time limit. 
In addition, the PDF of the surviving polymers tends to the
stable form 
\begin{equation}
\frac{p(s,t|s_0,N)}{S(t|s_0,N)}
\simeq
A_1(s_0,N)
\;X_1(s,N).
\end{equation}
Recent simulations 
\cite{Kardar4}
pointed out the long-time stability of $p(s,t|s_0,N)/S(t|s_0,N)$   
and revealed a singular behavior at the boundaries as a
specific anomalous signature of the translocation process 
\citep[See also][]{Kardar6}. 
By using the Frobenius method \citep[See, e.g.,][]{kusse} in Eq. (\ref{eq_fp_2}),
it can be proved that 
$X_1(s)\sim s^\phi$ for $s\to 0$ and 
$X_1(s)\sim (N-s)^\phi$ for $s\to N$, with 
\begin{equation}
\label{eq_phi}
\phi=\sigma+1
=(1+2\nu)\;(1-\alpha)+1.
\end{equation}
Such a singular behavior is only due to the divergence of $D(s,N)$ and
does not depend on the entropic drive. 
Putting $\alpha=0.81$ and $\nu=3/4$ in Eq. (\ref{eq_phi}), 
we thus get $\phi=1.52$, 
consistent with the simulation data (Fig. \ref{fig_pdf}c) and
very close to the value $\phi=1.44$ numerically found in
Ref. \cite{Kardar4}.
Using standard methods \cite{Risken1}, from Eq. (\ref{eq_fp_2}) one can also
deduce ordinary differential equations for the survival probability
$S(t|s_0,N)$ or for the mean translocation time 
$T(s_0,N)=\int_0^\infty d t\;S(t|s_0,N)$.

The theory performs extremely well for $s_0=N/2$.
Rescaling $s\mapsto s/N$ and $t\mapsto t/N^{\sigma+2}$ in
Eq. (\ref{eq_fp_2}), it is easily seen that the equation becomes 
$N$-independent as $N\gg1$. 
This implies that the  mean translocation
time scales as
\begin{equation}
\label{eq_scaling_time}
T(N/2,N)\sim N^{\sigma+2}.
\end{equation}
With $\alpha\simeq0.81$ for our 2d benchmark case, 
$\sigma+2\simeq2.48$, which is consistent
with what observed in our simulations and also in previous studies
\cite{Kardar1,Luo2}.
In the Supplemental Material, besides veryfying the validity of 
Eq. (\ref{eq_scaling_time}), we show a very satisfactory quantitative
comparison of the theoretical $T(N/2,N)$ with the one estimated from
the simulations. 
We stress, however, that different values of $\alpha$ lead to different
scaling exponents for the mean translocation time. Taking
$\alpha=2/(1+2\nu)$ gives $T(N/2,N)\sim N^{1+2\nu}$ in agreement
with~\cite{Kardar1}; 
Taking $\alpha=(1+\nu)/(1+2\nu)$ gives
$T(N/2,N)\sim N^{2+\nu}$ in agreement with~\cite{Barkema2}.
This is an important versatility of the theory, especially in view of
the fact that $\alpha$ has been recently found to be
viscosity-dependent \cite{Slater2}.
Fig. \ref{fig_survival} (a),(b) display a very good agreement of the
theory also with the numerical simulations of the process in terms of the
survival and translocation probability, respectively.
In particular, Fig. \ref{fig_survival} (b)
highlights the exponential decay of $Q(t|N/2,N)$ for large $t$.
More stringently, even $p(s,t|N/2,N)$ is accurately reproduced by
the theory for $t>\tau(N/2,N)$, as depicted in Fig. \ref{fig_pdf}
(a). In the long-time limit, the PDF of the surviving polymers
collapses onto the predicted form $A_1(N/2,N)\;X_1(s,N)$, inclusive of the singular
behavior at the borders [See Fig. \ref{fig_pdf} (b),(c)].

By considering $s_0$ closer to borders of the chain, the condition 
$S(\tau(N/2,N)|s_0,N)\simeq 1$ is violated.
According to our arguments above, 
the analysis of the mean translocation time
$T(s_0,N)$ as a function of $s_0$ reported in the Supplemental Material 
shows that indeed with $s_0$ becoming closer to the borders the mean translocation time
becomes slightly smaller than that predicted by our effective theory. 

We have shown that the late stages of unbiased polymer translocation
can be described 
by an effective Fokker-Planck  
equation whose solution
precisely reproduces the translocation PDF $p(s,t|s_0,N)$.
The finite-size scaling characterizing the Fokker-Planck diffusion
coefficient establishes a link with the anomalous diffusion at early
times, providing a convincing explanation of all the features of 
$p(s,t|s_0,N)$, including the singular behavior at
the border.

{\bf Acknowledgments}
This work is supported by ``Fondazione Cassa di Risparmio di Padova e Rovigo'' within the 
2008-2009 ``Progetti di Eccellenza'' program.

\bibliography{bibliography}

\begin{thebibliography}{25}
\bibitem[{foo({\natexlab{a}})}]{foot1}
\bibinfo{note}{$\tau(s_0,N)$, not be confused with the mean translocation time,
  has been practically defined as the time at which we observe a variation of
  5\% with respect to the extrapolation of the initial anomalous behavior of
  $\Delta s^2(t)$.}
\bibitem[{foo({\natexlab{b}})}]{foot2}
\bibinfo{note}{We use here It\^o's prescriptions for multiplicative noise, in
  place of Stratonovich's ones; The latter being frequently used in classical
  transport equations. We have verified that this choice does not affect
  significantly our results.}
\end{thebibliography}

\end{document}


\title{Finite-size scaling in unbiased translocation dynamics: Supplemental Material}

\author{Giovanni Brandani}
\affiliation{
University of Edinburgh, SUPA, School of Physics, Mayfield Road, EH9 3JZ, UK
}

\author{Fulvio Baldovin}
\affiliation{
Dipartimento di Fisica e Astronomia and Sezione INFN,
Universit\`a di Padova,
 Via Marzolo 8, I-35131 Padova, Italy
}

\author{Enzo Orlandini}
\affiliation{
Dipartimento di Fisica e Astronomia and Sezione INFN,
Universit\`a di Padova,
 Via Marzolo 8, I-35131 Padova, Italy
}

\author{Attilio L. Stella}
\affiliation{
Dipartimento di Fisica e Astronomia and Sezione INFN, 
Universit\`a di Padova,
 Via Marzolo 8, I-35131 Padova, Italy
}

\date{\today}

\begin{abstract}
We provide some technical derivations and further details.
\end{abstract}

\maketitle

\begin{figure}
\includegraphics[width=0.7\columnwidth]{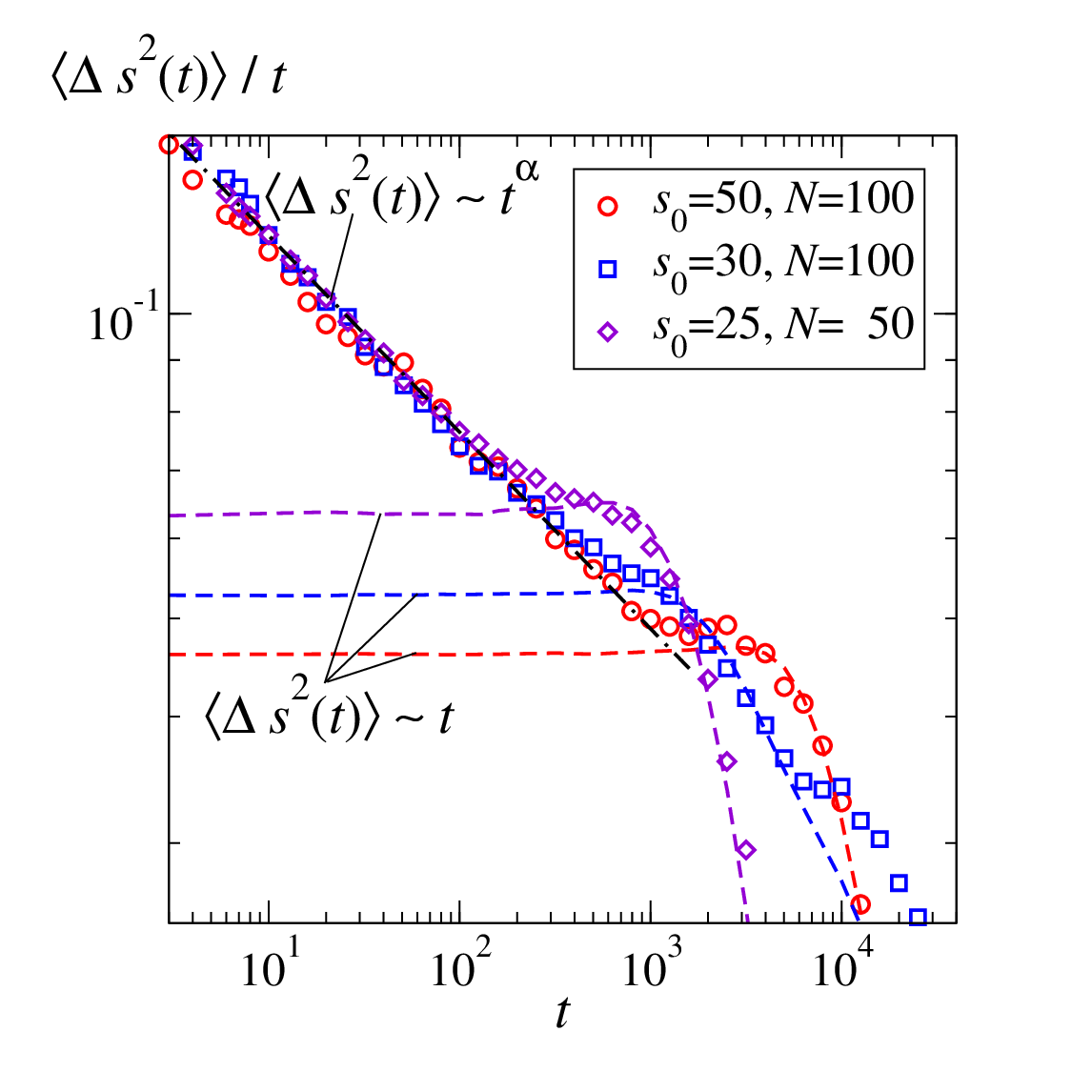}\\
\caption{
  Symbols: Time-rescaled mean square displacement evolution for a translocating
  chain. At early stages, the dynamics follows a
  universal sub-diffusive behavior
  $\langle\Delta s^2(t)\rangle\sim t^\alpha$, with
  $\alpha\simeq0.81$ (dot-dashed line).
  This regime breaks-down at a time $\tau$ depending on
  both $s_0$ and $N$. The dashed curves 
  are obtained by numerically solving Eq. (\ref{eq_fp_1}) 
  with $D(s,N)$ given by Eq. (\ref{eq_d_s}), $A=0.13$, and $s(0)=s_0$. 
}
\label{fig_msd_langevin}
\end{figure}

\section{Further numerical confirmations and limits of validity of the theory}
We comment evidence about the validity of our
theory and corresponding limitations,
based on Fig. 1 of the Main Text. For the Reader's convenience,
the same plots are reproduced in this Supplemental Material. 

Besides the results presented in Fig. 3, 4 of the Main Text, 
a further validation of our approach and an identification of its expecetd limit of validity comes from the 
comparison, given in Fig. \ref{fig_msd_langevin},
of the simulation results  (symbols) 
with the numerical solutions of the Fokker-Planck equation 
\begin{equation}
\label{eq_fp_1}
\partial_t\;p(s,t|s_0,N)=\partial^2_s\left[D(s,N)\;p(s,t|s_0,N)\right]
\end{equation}
with $D(s,N)$ given by 
\begin{equation}
\label{eq_d_s}
D(s,N)
=\frac{A}{N^\sigma}
\;\left[
\frac{1}{\left(\frac{s}{N}\right)^{1+2\nu}}
+\frac{1}{\left(1-\frac{s}{N}\right)^{1+2\nu}}
\right]^{1-\alpha},
\end{equation}
and initial condition
$p(s,0|s_0,N)=\delta(s-s_0)$ (dashed curves).

By comparison with Fig. 3a of the Main Text, dashed curves in Fig. \ref{fig_msd_langevin}
confirms that the approximation reported in the Main Text, 
\begin{eqnarray}
\partial_t\mathbb E\left[(s-s_0)^2\right]
&=&\int_0^N d s\;2\;D(s,N)\;p(s,t|s_0,N)
\nonumber\\
\label{eq_approximation}
&\simeq& 2\;D(s_0,N),
\end{eqnarray}
is justified up to time $t$ such that $S(t|s_0,N)\simeq1$. 
In addition, with $s_0=N/2$ simulations of Eq. (\ref{eq_fp_1})
are shown to 
well reproduce the time evolution of the mean square displacement
for $t>\tau(s_0,N)$, even when the effect of the absorbing boundaries
becomes important and $S(t|s_0,N)<1$. 
If we choose $s_0$ closer to the borders of the chain, deviations
appear in the long-time behavior
(See the $s_0=30$, $N=100$ plot). 

In our effective Fokker-Planck equation, different chemical
coordinates are entailed with different diffusion coefficients
$D(s,N)$. 
As can be appreciated through Fig. \ref{fig_msd_langevin} and
Eq. (\ref{eq_approximation}), 
during the initial
anomalous stage, the mean square displacement of the translocation
coordinate is larger than that effectively assumed
in the Fokker-Planck description.
Only for $t$ larger than the maximum $\tau$, i.e.,
$t>\tau(N/2,N)$, the anomalous stage may be considered ended and 
the effective diffusion coefficients $D(s,N)$ could be a
faithful representation of the ``diffusivity'' of the translocation
coordinate for all $s$.  
This poses a first limitation to our theory.
In addition, since we put as initial condition
$p(s,0|s_0,N)=\delta(s-s_0)$ in place of 
the PDF $p(s,\tau(N/2,N)|s_0,N)$ resulting from the initial anomalous
stage, 
we also need to assume that at $t=\tau(N/2,N)$ 
only a negligible fraction of polymers have
completed the translocation process, 
i.e., that $S(\tau(N/2,N)|s_0,N)\simeq1$. This condition is satisfied
for $s_0$ close to $N/2$. 
In view of the lower ``diffusivity'' in our representation, 
as $s_0$ becomes close to the borders we expect our
approach to provide an upper bound for the survival probability of a
translocating polymer. 
This is illustrated in Fig. \ref{fig_transl_time} of the Supplemental Material (See below).

\section{Derivation of the mean translocation time}
Our Fokker-Planck equation, Eq. (8) of the main text, may be rewritten
as 
\begin{equation}
\partial_t p(s,t|s_0,N)
=\mathcal L_s p(s,t|s_0,N),
\end{equation}
where the Liouvillian operator $\mathcal L_s$ is given by
\begin{equation}
\mathcal L_s
\equiv
\left\{
\partial_s\;D(s,N)\;\frac{\partial\ln\Omega(s,N)}{\partial s}
\right\}
+\left\{
\partial^2_s\;D(s,N)
\right\}.
\end{equation}

Defining the survival probability $S(t|s_0,N)$ as
\begin{equation}
S(t|s_0,N)\equiv\int_0^N ds\;p(s,t|s_0,N),
\end{equation}
it follows \cite{Risken1} that $S(t|s_0,N)$ satisfies the equation:
\begin{equation}
\label{eq_surv}
\partial_t S(t|s_0,N)
=\mathcal L_{s_0}^\dag S(t|s_0,N), 
\end{equation}
where the adjoint Liouvillian operator $\mathcal L_{s_0}^\dag$ (acting now
on the initial data $s_0$) is
\begin{equation}
\mathcal L_{s_0}^\dag
=
\left\{
D(s_0,N)\;\frac{\partial\ln\Omega(s_0,N)}{\partial s_0}\;\partial_{s_0}
\right\}
+
\left\{
D(s_0,N)\;\partial_{s_0}^2
\right\}
.
\end{equation} 

In terms of $S(t|s_0,N)$ the mean translocation time $T(s_0,N)$ is
given by
\begin{eqnarray}
T(s_0,N)&=&-\int_0^\infty dt \;t\;\partial_tS(t|s_0,N)\\
&=&\int_0^\infty dt \;S(t|s_0,N),
\end{eqnarray}
inheriting from Eq. (\ref{eq_surv}) the ordinary differential equation
\begin{equation}
\label{eq_mtt}
\mathcal L_{s_0}^\dag T(s_0,N)=-1.
\end{equation}

\begin{figure}
\includegraphics[width=0.7\textwidth]{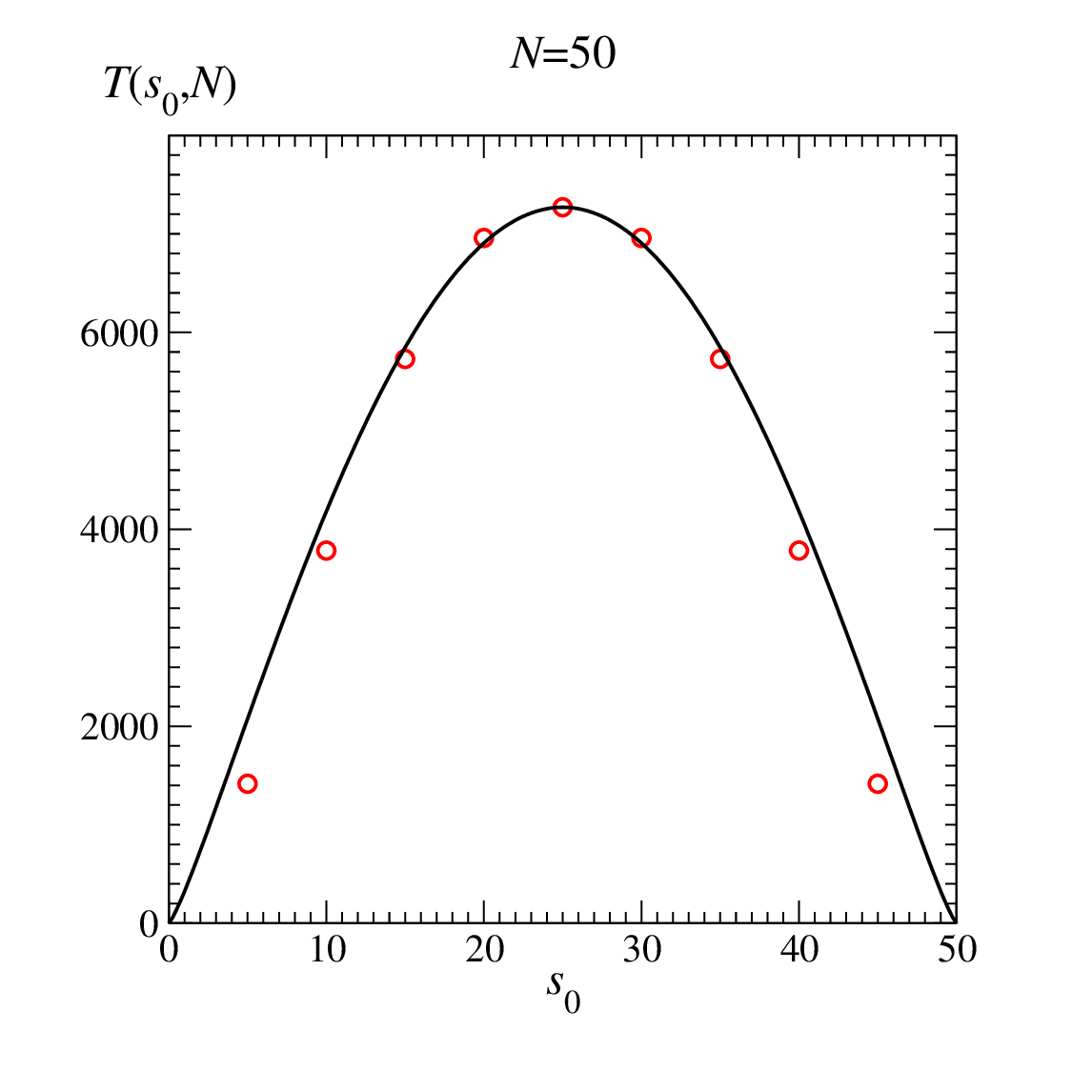}
\caption{
  Mean translocation time as a function of $s_0$:
  MD simulations (symbols) vs. theory (full lines).
}
\label{fig_transl_time}
\end{figure}

Eq. (\ref{eq_mtt}) has boundary conditions 
\begin{eqnarray}
T(0,N)&=&0,
\\
T(N,N)&=&0,
\end{eqnarray}
and can be solved, e.g., numerically for a generic $s_0$. 
Numerical solutions 
for $N=50$ are compared in Fig. \ref{fig_transl_time} with the results
of molecular dynamics simulations of the translocation dynamics as
described in the Main Text. 
Taking advantage of the symmetry with respect to the middle point,
here we explicitly calculate the mean translocation time for
$s_0=N/2$.
We first rewrite Eq. (\ref{eq_mtt}) as
\begin{equation}
\left\{
\partial_{s_0}\;\Omega(s_0,N)\;\partial_{s_0}
\right\}
\;T(s_0,N)
=
-\frac{\Omega(s_0,N)}{D(s_0,N)}.
\end{equation}
Integrating between $N/2$ and $s_0$ we have
\begin{equation}
\label{eq_step_1}
\Omega(s_0,N)\;\partial_{s_0}T(s_0,N)
=
-\int_{N/2}^{s_0}ds_0^\prime\;\frac{\Omega(s_0^\prime,N)}{D(s_0^\prime,N)},
\end{equation}
where we have used that
$\displaystyle\left.\partial_{s_0}T(N/2,N)\right|_{s_0=N/2}=0$ as
implied by the symmetry of the problem around $s_0=N/2$.  
Rearranging Eq. (\ref{eq_step_1}) and integrating between $N/2$
and $N$ we finally obtain
\begin{equation}
\label{eq_mtt_2}
T(N/2,N)
=\int_{N/2}^Nds_0\;\frac{1}{\Omega(s_0,N)}
\;\int_{N/2}^{s_0}ds_0^\prime\;\frac{\Omega(s_0^\prime,N)}{D(s_0^\prime,N)},
\end{equation}
where we have used the boundary condition $T(N,N)=0$.

\begin{figure}
\includegraphics[width=0.7\textwidth]{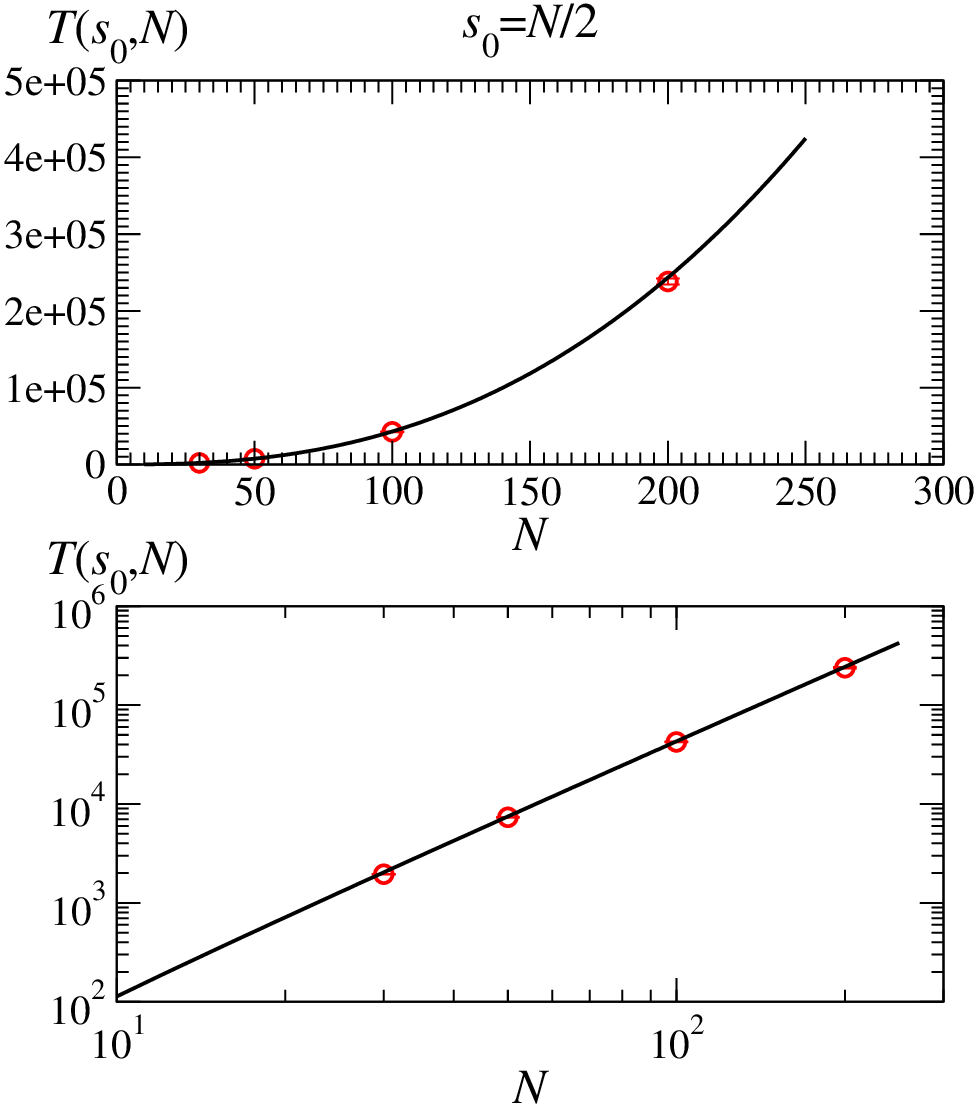}
\caption{
  $N$-dependence of the mean translocation time at $s_0 = N/2$ in linear and log-log scales:
  Symbols refer to MD simulations while the full curves refer to Eq. (\ref{scaling}) 
  with $A=0.13$, $b_0=0.4$,  and $\sigma = (1+2\nu)(1-\alpha)$ with $\nu = 3/4$, $\alpha = 0.81$.
}
\label{fig_transl_time_scaling}
\end{figure}

Eq. (\ref{eq_mtt_2}) can be evaluated explicitly by using the
expressions for $D(s_0,N)$ and $\Omega(s_0,N)$ given in the Main
Text. Namely,
\begin{equation}
D(s_0,N)
=\frac{A}{N^\sigma}
\;\left[
\frac{1}{\left(\frac{s_0}{N}\right)^{1+2\nu}}
+\frac{1}{\left(1-\frac{s_0}{N}\right)^{1+2\nu}}
\right]^{1-\alpha},
\end{equation}
\begin{equation}
\Omega(s_0,N)
\propto
[s_0\;(N-s_0)]^{(\gamma_1-1)}
\;C(s_0)\;C(N-s_0),
\end{equation}
with
\begin{equation}
C(s_0)\simeq1+\frac{b_0}{s_0^{1/2}}.
\end{equation}
The result is
\begin{eqnarray}
T(N/2,N)
&=&
\frac{N^{\sigma+2}}{A}
\;\int_{1/2}^1dx
\frac{1}{
[x\;(1-x)]^{(\gamma_1-1)}
\;\left[1+\frac{b_0}{N^{1/2}\;x^{1/2}}\right]
\;\left[1+\frac{b_0}{N^{1/2}\;(1-x)^{1/2}}\right]
}
\;\cdot
\nonumber\\
&&
\;\cdot
\;\int_{1/2}^1dx^\prime
\frac{
[x^\prime\;(1-x^\prime)]^{(\gamma_1-1)}
\;\left[1+\frac{b_0}{N^{1/2}\;{x^\prime}^{1/2}}\right]
\;\left[1+\frac{b_0}{N^{1/2}\;(1-x^\prime)^{1/2}}\right]
}{
\;\left[
\frac{1}{{x^\prime}^{1+2\nu}}
+\frac{1}{\left(1-x^\prime\right)^{1+2\nu}}
\right]^{1-\alpha},
\label{scaling}
}
\end{eqnarray}
where $\sigma=(1+2\nu)\;(1-\alpha)$.
As $N\gg1$ the correction to scaling can be neglected, and the above
expression simplfies into 
\begin{eqnarray}
T(N/2,N)
&=&
\frac{N^{\sigma+2}}{A}
\;\int_{1/2}^1dx
\frac{1}{
[x\;(1-x)]^{(\gamma_1-1)}
}
\;\int_{1/2}^1dx^\prime
\frac{
[x^\prime\;(1-x^\prime)]^{(\gamma_1-1)}
}{
\;\left[
\frac{1}{{x^\prime}^{1+2\nu}}
+\frac{1}{\left(1-x^\prime\right)^{1+2\nu}}
\right]^{1-\alpha}
}
\nonumber\\
&\propto&
N^{\sigma+2},
\label{scaling_2}
\end{eqnarray}
yielding the expected scaling behavior. 
In Fig.~\ref{fig_transl_time_scaling} we compare the analytical expression  (\ref{scaling}) 
with numerical data based on MD simulations:  
The agreement turns out to be  extremely good.

\bibliography{bibliography}